# Motion of an active particle in a linear concentration gradient

Prathmesh M. Vinze, Akash Choudhary, S. Pushpavanam[*]

Department of Chemical Engineering, Indian Institute of Technology Madras,
Chennai, TN 600036, India

Janus particles self-propel by generating local tangential concentration gradients along their surface. These gradients are present in a layer whose thickness is small compared to the particle size. Chemical asymmetry along the surface is a pre requisite to generate tangential chemical gradients, which gives rise to diffusioosmotic flows in a thin region around the particle.This results in an effective slip on the particle surface. This slip results in the observed "swimming" motion of a freely suspended particle even in the absence of externally imposed concentration gradients. Motivated by the chemotactic behavior of their biological counterparts (such as sperm cells, neutrophils, macrophages, bacteria etc.), which sense and respond to external chemical gradients, the current work aims at developing a theoretical framework to study the motion of a Janus particle in an externally imposed linear concentration gradient. The external gradient along with the self-generated concentration gradient determines the swimming velocity and orientation of the particle. The dominance of each of these effects is characterized by a non-dimensional activity number $A$ (ratio of applied gradient to self-generated gradient). The surface of Janus particle is modelled as having a different activity and mobility coefficient on the two halves. Using the Lorentz Reciprocal theorem, an analytical expression for the rotational and translational velocity is obtained. The analytical framework helps us divide the parameter space of surface activity and mobility into four regions where the particle exhibits different trajectories.

[*]E-mail: spush@iitm.ac.in

# 1. Introduction

Artificial active particles can swim autonomously in a fluid even in the absence of an external electrical or chemical field. These particles convert the chemical energy of the solute molecules into mechanical energy; their *asymmetric nature* provides a sense of directionality/navigation. The self-propulsion of these micron to submicron sized particles falls in the regime of low Reynolds numbers (~$10^{-6}$), where symmetry breaking is essential for swimming[1]. This is realised by introducing an asymmetry in the surface properties such as surface activity (adsorption, desorption, or reaction), surface mobility (molecular interactions with particle surface)[2,3]. This generates near-surface tangential gradients in potential energy necessary for self-propulsion. Since these particles mimic swimming of microorganisms, their study helps develop insights into potential targeted drug delivery techniques in medical therapy[4]. Many microorganisms exhibit chemotaxis i.e., response to chemical signals present in their environment. For instance, chemical signals sent out by mammalian eggs help sperm cells to *find* them[5,6]. Bacteria such as *E.coli* moves towards nutrients such as ribose and galactose while running away from phenol by temporally sensing the chemical gradient and accordingly regulating its complex flagellar rotations [7–9]. Recent works on artificial swimmers has shown significant similarity between artificial and biological chemotaxis[10,11]. Artificial swimmers can 'seek' out a target by sensing chemical gradients generated by the diseased/infectious site similar to their biological counterparts and deliver medicinal payloads[4,12]. Most of the effort so far has focused on simulating active particle in a uniform concentration of an inert particle in a concentration gradient.

Derjaguin[13] and co-workers were one of the first researchers to study diffusiophoresis: movement of colloidal particles in concentration gradient. Later, using a continuum framework, Anderson[14] worked on diffusioosmosis at the surface of a freely suspended inert particle, which results in its movement. Here an inert particle moves towards a higher or lower concentration region based on its interaction with the solute molecules. The interaction between solute molecules and the particle is restricted to a thin layer, giving rise to a pressure gradient. This drives the fluid inside the thin layer, which can be viewed as a slip at the surface. Experimentalists have recently synthesized rod-shaped self-electrophoretic particles where the two halves of the particles act as sites for redox reactions[15–17]. This creates a local ionic gradient resulting in an electro-osmotic slip at the surface; in turn generating a particle movement called 'self-electrophoresis'. Golestanian et.al[2,18] studied another mechanism of autonomous swimming. They analysed a particle which is coated with platinum on one half of its surface and is coated with non-conducting polystyrene inert on the other half, placed in a uniform concentration of hydrogen peroxide. The solute molecules react/adsorb on the 'active side' and create a local concentration gradient along the surface. This induces a diffusio-osmotic flow which results in 'self-diffusiophoresis' of a freely suspended particle. Golestanian et. al[3] provided a generalized framework for self-diffusiophoresis and self-electrophoresis and showed that these two swimming mechanisms are analogous to each other[3,17]. On the basis of existing experimental studies three primary assumptions were made: 1) the interactive layer is thin compared to the size of the particle. This helped them carry out an asymptotic analysis and enabled replacing the diffusio-osmotic flow of the thin interactive layer with a slip at the surface; 2) solute transport occurred primarily via diffusion i.e. advective effects were negligible; 3) a fixed rate of adsorption/desorption of solute capture/release at the active sites. Their formulation showed that a chemically active particle like a Janus sphere required symmetry breaking in activity for self-propulsion.

Khair[19] extended the work of Anderson et.al[14] to include the effect of solute advection. Using a perturbation expansion in Peclet number (ratio of advective to diffusive effects), the effect of solute advection on phoretic swimming was explored. The phoretic translation velocity was found to be monotonically decreasing with increasing Pe. For a slightly non-spherical particle, the translating velocity was found to be dependent on shape and orientation. This was in contrast to the case where velocity is independent of size and shape for Pe=0.

Table 1: Summary of earlier theoretical studies on diffusiophoresis and self-diffusiophoresis.

| Investigation | Regime | Remarks/Description |
|---|---|---|
| **Anderson et.al (1982)** | $Pe \to 0, Re \to 0$ | Diffusiophoretic swimming of an inert particle in an external concentration gradient. |
| **Golestanian et.al (2007)** | $Pe \to 0, Re \to 0, \frac{\delta}{a} \to 0$ | Provided a unified framework describing phoretic swimming using foundations laid by Derjaguin (1947) and Anderson et.al[14] |
| **Khair (2014)** | $Pe \geq O(1), Re \to 0$ | Extended the work of Anderson et. al[14] to include convective effects as higher order effects in $Pe$. |
| **Popescu et.al (2018)** | | A qualitative study of chemotaxis of a Janus sphere. |

| This work | $Pe \to 0, Re \to 0, \dfrac{\delta}{a} \to 0$ | Provides a framework for quantitative understanding of artificial chemotaxis. |
|---|---|---|

Taking inspiration from immune cells (such as neutrophils and macrophages) that respond to chemical gradients and move towards the site of injury/ infection[20], the current work aims at analysing the response of an artificial swimmers to an external concentration gradient. Very recently, Popescu et.al[12] qualitatively analysed the response of a Janus sphere placed in a concentration gradient and showed that the reorientation of the Janus sphere along the direction of concentration gradient requires an additional symmetry breaking in the solute surface interaction i.e., a gradient in surface mobility. They showed that particle movement can be generated by two different effects i.e. an externally imposed concentration gradient and the self-generated concentration gradient.

Motivated by the need to provide guidelines for rational fabrication of drug delivery systems, the quantitative chemotactic response and its dependence on different parameters must be analysed. It is hence important to seek answers to questions such as (i) what is the time required for reorientation? (ii) how does this re-orientation compare with that induced by Brownian noise? (iii) how does the swimming direction depend on the relative strength of the self-generated and the artificially imposed concentration gradients? (iv) how does the trajectory depend on surface activity, mobility coefficient.

To obtain insights into the motion of biological swimmers and keeping applications such as drug delivery in perspective, in this work, we theoretically study the motion of an active particle placed in an external concentration gradient. The interaction between the external concentration gradient with the self-generated concentration gradient is characterised by a dimensionless activity number $A$ which is the ratio of external concentration gradient to self-generated concentration gradient. We discuss the various characteristic scales governing the system and state the governing and boundary equations in section 2. In section 3, using the principle of linearity, the governing equations are decomposed into two subproblems: the first takes into account the external concentration gradient while the second accounts for the surface activity. The next section (4) is dedicated to finding the swimming and rotational velocity using the Lorentz reciprocal theorem. Using the derived translational and rotational velocity, dynamic equations are used to determine the trajectory of the particle. In section 5, we quantify the rotational and translational behaviour of the particle. Finally, in section 6, we discuss the key insights and conclusions.

## 2. Problem formulation

A Janus particle propels itself in a solution with uniform external concentration by creating a local tangential concentration gradient. The solute molecules interact with the surface of the Janus particle in a thin region. The interactions of solute molecules with the particle are characterised by a mobility coefficient. A positive mobility coefficient refers to repulsive interactions, while a negative value refers to attractive interactions. Through the concentration gradient, the interaction generates a pressure gradient, which drives the diffusion-osmotic flow inside the thin layer. The diffusioosmotic flow acts as an effective slip and results in a swimming motion of the Janus particle. In this work, we consider a particle which has two different axisymmetric surface activities (shown by red and blue colour in Fig.1) on the two halves; similarly, the two faces have different mobility coefficients. When placed in a fluid with uniform concentration of solute molecules, a Janus particle swims along the axis of symmetry, also called axis of self-propulsion ($e_z$ in Fig.1a).

We study the motion of a Janus particle of radius '$a^*$' under the influence of an external linear concentration gradient of strength $\gamma^*$, where the superscript * is used to represent dimensional variables. We define two frames of reference centered on the particle: a frame of reference ($\boldsymbol{e}_y, \boldsymbol{e}_z$) whose z- axis coincides with the axis of self-propulsion at every instant of time, and a stationary frame of reference given by ($\boldsymbol{e}_{y_0}, \boldsymbol{e}_{z_0}$). The externally imposed concentration gradient is at an angle $\beta_0$ with horizontal axis, which is denoted by $e_{z_0}$ in a stationary frame of reference. Fig.1a shows the position of the particle and the axis in the particle reference frame at t=0. Fig. 1b shows the relationship between the two different reference frames at a later instant of time. The instantaneous angle between the concentration gradient and its axis of self-propulsion ($e_z$) is given by $\beta(t)$ (as shown in Fig 1a). The co-rotational frame of reference is used to find translational and rotational velocity, while the particle trajectory is tracked in the stationary reference frame. At $t = 0$ (i.e., $\beta = \beta_0$) the two frames of reference coincide (assuming the particle to be horizontal initially).

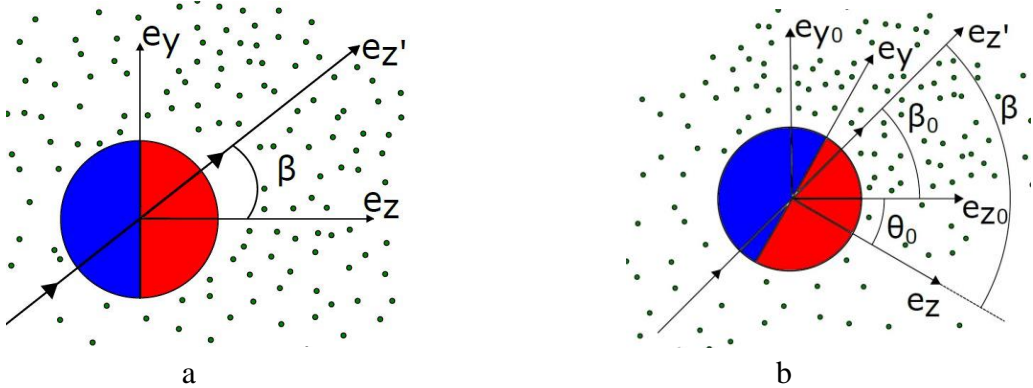

Figure 1: Schematic elucidating the reference frames used in this work. The external gradient makes a constant angle $\beta_0$ with respect to $\boldsymbol{e}_{z_0}$ (non-rotational frame). The instantaneous angle between the axis of self-propulsion and concentration gradient is given by $\beta$. The two halves have different activity and mobility coefficients represented by diferent colour on the two halves. Mobility and activity on the red half is represented by a '+' subscript, whereas '-' subscript is used for the blue half. $e_z$ and $e_y$ represent the particle co-rotational frame of reference and $e_{y_0}$, $e_{z_0}$ represents the stationary frame of reference in Fig 1b. The direction of concentration gradient is shown by $e_{z'}$.

Activity on the surface $\alpha^*(\theta, \phi)$ represents consumption or release of solute on the particle surface, and is expressed as

$$\alpha^* = \begin{cases} \alpha_+^* & 0 < \theta < \dfrac{\pi}{2} \\ \alpha_-^* & \dfrac{\pi}{2} < \theta < \pi \end{cases}$$

Here, $\alpha_+^*$ and $\alpha_-^*$ are the activities on the two faces, and $\theta$ is the polar angle. The activity is uniform in the axisymmetric direction. $\boldsymbol{e_{z'}}$ is the direction of external concentration gradient. Here the concentration field is of the form $C_\infty^* = \gamma^* z'$; where $z'$ is the distance of a point on $\boldsymbol{e_{z'}}$ from the origin (as shown in Fig 1a ). $z'$ is expressed in terms of $y, z$ coordinates. For this we take the projection of $y$ and $z$ coordinates along the $\boldsymbol{e_{z'}}$ axis. This yields, $z' = z\cos\beta + y\sin\beta$. The external concentration field can be expressed as, $C_\infty^* = \gamma^*(z\cos\beta + y\sin\beta)$. Using $z = r\cos\theta$ and $y = r\sin\theta\sin\phi$, $\theta$ and $\phi$ being the polar and radial angles, we obtain:

$$C_\infty^* = \gamma^* r\left(\cos\theta\cos\beta + \sin\theta\sin\phi\sin\beta\right) \tag{1}$$

The solute and momentum transport around the particle is characterised by the following timescales:

i) momentum diffusion $t_{mom} = a^{*2}/\nu^* \sim O(10^{-4})$s
ii) solute diffusion $t_{diff} = a^{*2}/D^* \sim O(10^{-1})$s
iii) particle translation $t_{swim} = a^*/u_{ch}^*$
iv) particle rotationa $t_{rot} = a^*/\mu^*\gamma^*$.

Here, $D^*$ is the solute diffusion coefficient, $\mu^*$ is the mobility coefficient, $\nu^*$ is the kinematic viscosity.

The longest timescale (slowest process) characterises the temporal change in solute and momentum transport. The estimates of the different time scales were obtained by considering a Janus particle of size $10\mu m$ in a concentration gradient of $10\ Mm^{-1}$; the diffusion coefficient of oxygen in water $= 2.3 \times 10^{-9} m^2 s^{-1}$; kinematic viscosity of water $= 1.05 \times 10^{-6} m^2 s^{-1}$. The mobility coefficient for the Janus particle with oxygen gas as a solute is estimated as[21] $\mu = \dfrac{k_B T d^{*2}}{2\eta^*} = 8.87 \times 10^{-33} m^5 s^{-1}$; For a swimming velocity of the order ~1µm/s, the various timescales are found to be related as

$$t_{diff} \ll t_{rot} \tag{2}$$

$$t_{diff} \ll t_{swim} \tag{3}$$

The timescale of diffusion being lower than those of both swimming and rotation suggests that the concentration field is primarily determined by the particle orientation. This allows us to make pseudo steady state approximation i.e. we can neglect the unsteady terms in the governing equations. The characteristic velocity scale $u_{ch}^*$ has two contributions: one arising from the externally imposed concentration gradient (scaled as $\gamma^*\mu^*$), and another arising from the activity generated concentration gradients (scaled as $\frac{|\alpha^*|\mu^*}{D^*}$), here $|\alpha^*|$ represents the maximum magnitude of the surface activity. To depict the ratio of two contributions, we define a dimensionless activity number as $A = \frac{\gamma^* D^*}{|\alpha^*|}$. We chose the velocity scale as $u_{ch}^* = \mu^*\left(\gamma^* + \frac{|\alpha^*|}{D^*}\right) = \frac{|\alpha^*|\mu^*}{D^*}(A+1)$, this takes into account both the contributions. We see, for $A \ll 1$, $u_{ch}^* \sim \frac{|\alpha^*|\mu^*}{D^*}$ and for $A \gg 1$, $u_{ch}^* \sim \gamma^*\mu^*$.

We now seek bounds on external concentration gradient and the surface activity for the quasi-steady approximation to be valid. For this we substitute the expression of $t_{diff}$, $t_{swim}$ and $t_{rot}$ in equations (2) and (3)

$$\frac{t_{diff}}{t_{rot}} = \frac{a^* \gamma^* \mu^*}{D^*} \ll 1 \Rightarrow \gamma^* \ll \frac{D^*}{a^* \mu^*}$$

$$\frac{t_{diff}}{t_{swim}} = \frac{|\alpha^*| a^* \mu^*}{D^{*2}} \ll 1 \Rightarrow |\alpha^*| \ll \frac{D^{*2}}{a^* \mu^*} \quad (4)$$

Substituting the values from above, we obtain an upper bound on $\gamma^*$ and $|\alpha^*|$,
$$\gamma^* \ll 1.66 \times 10^3 \frac{M}{m} \text{ and } |\alpha^*| \ll 1.66 \times 10^{-3} M\, m^{-2} s^{-1}$$

Apart from the above bound, there is another upper bound on $\gamma^*$ which arises from the weak gradient condition. Under a strong external concentration gradient, the transport becomes unsteady[14]. To ensure pseudo steady-state the gradient has to be sufficiently weak. This yields,

$$\frac{\gamma^*}{c_0/a^*} \ll 1$$

A typical value for solute concentration $c_0$ is[19] 0.1 M, which yields $\gamma \ll 10^4$ M/m. This is satisfied when the bound for quasi-steady state approximation i.e. eq.(4) holds.

Choosing the characteristic concentration scale as $C_{ch} = \frac{|\alpha^*|a^*}{D^*}$, lengthscale $l_{ch} = a^*$, timescale $t_{ch} = t_{rot} = \frac{a^*}{\gamma^*\mu^*}$ and the velocity scale $u_{ch}^*$ as $\frac{|\alpha^*|\mu^*}{D^*}(A+1)$ the dimensionless solute balance and momentum balance equations are

$$Pe\left(\left(\frac{A}{A+1}\right)\frac{\partial c}{\partial t} + \nabla \cdot (\mathbf{u}c)\right) = \nabla^2 c \quad (5)$$

Subject to the flux condition
$$-\mathbf{n} \cdot \nabla c = \alpha(\theta) \quad (6)$$

And the far field condition
$$c \to C_\infty \quad \text{as } r \to \infty \quad (7)$$

The fluid velocity is governed by

$$\mathrm{Re}\left(\left(\frac{A}{A+1}\right)\frac{\partial \mathbf{u}}{\partial t}+\mathbf{u}\cdot\nabla\mathbf{u}\right)=-\nabla P+\nabla^2\mathbf{u} \qquad (8)$$

Subject to the boundary conditions

$$\mathbf{u}=\mathbf{u}_s+\mathbf{U}_{swim}+\Omega\times\mathbf{r} \quad \text{at } r=1 \qquad (9)$$

No penetration condition

$$\mathbf{n}\cdot\mathbf{u}=0 \quad \text{at } r=1 \qquad (10)$$

Far field condition

$$\mathbf{u}\to 0 \quad \text{as } r\to\infty. \qquad (11)$$

Here, the slip velocity $\boldsymbol{u}_s=\mu_0(\mathbf{I}-\mathbf{nn})\cdot\nabla c$; where $\mu_0$ is the scaled mobility coefficient, and $\mathbf{I}$ is the identity tensor. The equations (5)-(11) govern the solute and momentum transport and are expressed in the particle reference frame. Two dimensionless numbers describing the system behavior are $Pe=\frac{u^*_{ch}a^*}{D^*}$ and $Re=\frac{u^*_{ch}a^*}{\nu^*}$. For particle of size 10µm, with the swimmer velocity of the order ~ 1 µm/s, taking diffusion coefficient as $2.3\times 10^{-9} m^2/s$ and kinematic viscosity of water as $1.05\times 10^{-6}\ m^2/s$; the associated Reynolds number and Peclet number are $9.52\times 10^{-6}$ and $4.3\times 10^{-2}$ respectively. The low values of these dimensional numbers suggests that the inertial and advective terms can be neglected both in solute and momentum transport. In the limit $Pe\to 0$, the two equations (5) and (8) are decoupled. We first solve for the concentration field and obtain the slip velocity at the surface. The solute molecule interacts with the particle surface in a thin region. The thickness of this interaction layer ($\delta^*$) is of the order of a few nano meters[21] (Anderson 1989), resulting in $\frac{\delta^*}{a^*} \sim O(10^{-4})$. For such ratios, the fluid velocity profile inside this layer can be visualised as a slip at the surface. This slip velocity is employed in the boundary condition (9).

In the limit $Pe\to 0$, the solute balance equation is given by

$$\nabla^2 c=0 \qquad (12)$$

$$-\mathbf{n}\cdot\nabla c=\alpha(\theta) \quad \text{at } r=1 \qquad (13)$$

$$c\to C_\infty \quad \text{as } r\to\infty \qquad (14)$$

here $C_\infty = Ar(\cos\theta\cos\beta+\sin\theta\sin\phi\sin\beta)$. Introducing the disturbance concentration field as, $c'=c-C_\infty$ and substituting in the governing equations, we obtain:

$$\nabla^2(c'+C_\infty)=0 \qquad (15)$$

$$-\mathbf{n}\cdot\nabla(c'+C_\infty)=\alpha(\theta) \quad \text{at } r=1 \qquad (16)$$

$$c'\to 0 \quad \text{as } r\to\infty \qquad (17)$$

Since the external concentration field is linear, $\nabla^2 C_\infty=0$. The governing equation in terms of disturbance variable $c'$ is

$$\nabla^2 c'=0 \qquad (18)$$

Writing the equations in terms of disturbance field results in both the non homogeneities arising in the boundary condition at the particle surface. Simplifying (16) and (17) we obtain

$$-\mathbf{n}\cdot\nabla c' = A(\cos\theta\cos\beta + \sin\theta\sin\phi\sin\beta) + \alpha(\theta) \tag{19}$$

$$c' \to 0 \text{ as } r \to \infty \tag{20}$$

In the next section, we seek analytical solution for (18)-(20) exploiting the linearity of the above system.

## 3. Solution for concentration fields

The governing equations (18)-(20) are linear with two non-homogeneities in the boundaries: the first non-homogeneity arises due to the external concentration gradient (19), while the second one arises from the activity on the particle surface $\alpha(\theta)$ (19). Using the principle of superposition, the governing equations (18)-(20) are decomposed into two sub problems with one non-homogeneity each. We seek the solution for disturbance concentration as $c' = c_1 + c_2$. These sub-problems are governed by the following equations

$$\nabla^2 c_1 = 0 \tag{21}$$

$$-\mathbf{n}\cdot\nabla c_1 = A(\cos\theta\cos\beta + \sin\theta\sin\phi\sin\beta) \quad \text{at } r = 1 \tag{22}$$

$$c_1 \to 0 \quad \text{as } r \to \infty, \tag{23}$$

and

$$\nabla^2 c_2 = 0 \tag{24}$$

$$-\mathbf{n}\cdot\nabla c_2 = \alpha(\theta) \quad \text{at } r = 1 \tag{25}$$

$$c_2 \to 0 \quad \text{as } r \to \infty. \tag{26}$$

$c_1$ represents the disturbance concentration field of a *passive* diffusiophoretic sphere in an external linear concentration gradient. Whereas, the '$c_2$ problem' represents the disturbance concentration field around a Janus sphere.

### 3.1 Concentration field around a diffusiophoretic particle in linear concentration gradient

The solute concentration field of this sub problem governed by equations (21)-(23). The equations are linear in $\nabla C_\infty$. A permissible decaying solution that is linear in $\nabla C_\infty$ is

$$c_1 = B\frac{(\nabla C_\infty)\cdot(\mathbf{x})}{r^3}, \tag{27}$$

Where $x$ is the positional vector of a point, $B$ is a constant which will be determined by the surface boundary condition (22). Substituting $C_\infty = A(r\cos\theta\cos\beta + r\sin\theta\sin\phi\sin\beta)$, results in

$$c_1 = \frac{BA(\cos\theta\cos\beta + \sin\theta\sin\phi\sin\beta)}{r^2}. \tag{28}$$

The boundary condition (22) in spherical coordinates is

$$-\frac{dc_1}{dr}\bigg|_{r=1} = A(\cos\theta\cos\beta + \sin\theta\sin\phi\sin\beta) \tag{29}$$

Substituting $c_1$ from (28) we obtain $B = \frac{1}{2}$ and the solution for $c_1$ as

$$c_1 = \frac{A(\cos\theta\cos\beta + \sin\theta\sin\phi\sin\beta)}{2r^2} \tag{30}$$

The disturbance field in the y-z plane is plotted in the Fig 2a. The concentration gradient is along a line inclined at an angle $\pi/4$ with respect to $e_z$ axis ($\beta = \pi/4$). The disturbance field is symmetric about the direction of the concentration gradient(shown by the red dashed line in Fig 2a), and it decays as $r^{-2}$.

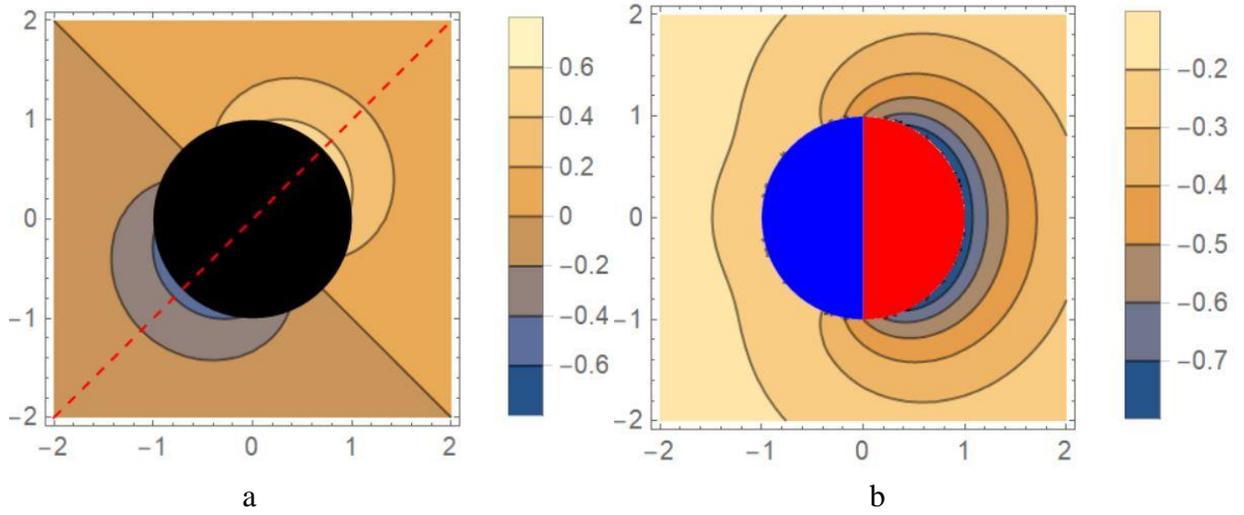

a                                                              b

Figure 2: Disturbance concentration field for (a) a passive diffusiophoretic particle in an external linear concentration gradient along $\beta = \pi/4$ and (b) Janus particle in a uniform concentration with x-axis as self propulstion axis . The concentration field for (a) decays as $O(r^{-2})$ whereas for (b) at the leading order it decays as $O(r^{-1})$. Here, the red face of the Janus sphere absorbs the solute molecules due to which there is a drop in concentration near the red face, $\alpha_+ = +1, \alpha_- = 0$.

## 3.2 Concentration field around a Janus particle

Following Golestanian et.al[3] (2008) the solution to this problem is sought using Legendre Polynomials as a basis set. This is given by

$$c_2 = \sum_{l=0}^{l=\infty} \frac{\alpha_l}{l+1} r^{-(l+1)} P_l(\cos\theta) \tag{31}$$

Here, $\alpha_l$ represents the coefficients of activity $\alpha(\theta)$ expanded in terms of Legendre polynomials, $\alpha(\theta) = \sum_{l=0}^{l=\infty} \alpha_l P_l(\cos\theta)$. The disturbance concentration field is shown in Fig 2b.

The composite disturbance field is given by $c' = c_1 + c_2$ is

$$c' = \frac{A(\cos\theta\cos\beta + \sin\theta\sin\phi\sin\beta)}{2r^2} + \sum_{l=0}^{l=\infty} \frac{\alpha_l}{l+1} r^{-(l+1)} P_l(\cos\theta) \qquad (32)$$

And the complete concentration field $c$ given by $c = c' + C_\infty$.

$$c = Ar(\cos\theta\cos\beta + \sin\phi\sin\theta\sin\beta)(1+\frac{1}{2r^3}) + \sum_{l=0}^{l=\infty} \frac{\alpha_l}{l+1} r^{-(l+1)} P_l(\cos\theta) \qquad (33)$$

Equation (33) indicates there are two contributions to the concentration field. For $A \gg 1$, the effect of the external concentration gradient is dominant, and the particle behaves similar to a passive particle placed in an external concentration gradient. For $A \ll 1$, the local concentration gradient generated by the asymmetric surface dominates, and the particle behavior is primarily governed by the active Janus particle. When $A \sim O(1)$, both the effects contribute equally. In Fig 3, concentration contours of a Janus particle in a linear concentration gradient for different values of A is shown for $\beta_0 = \pi/4$. Fig3a shows that the external gradient contribution is low, and the concentration field is symmetric about the x-axis for $A = 0.01$. For an intermediate value of $A = 0.1$, both the contributions are significant as shown in Fig. 3b. Whereas, Fig 3c shows the external gradient contribution dominates for $A = 10$.

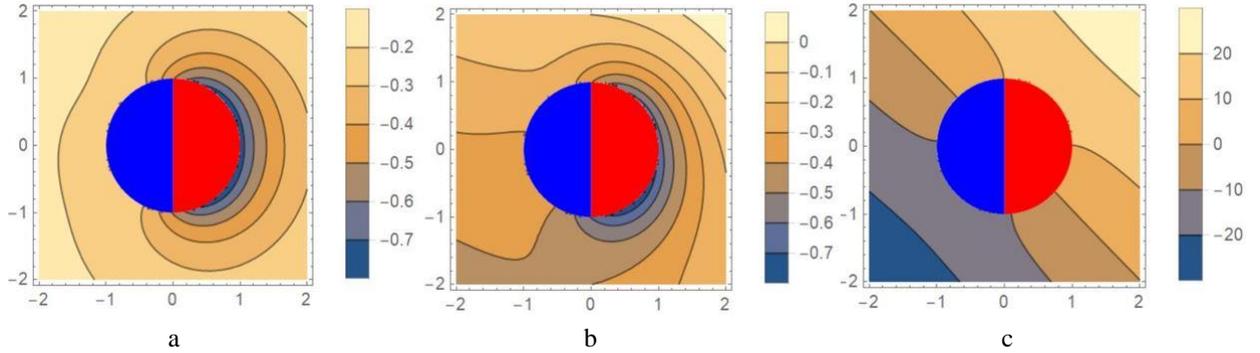

a  b  c

Figure 3: Concentration field around a Janus particle placed in a linear concentration gradient for different activity number a) 0.01 b) 0.1 c) 10 with $\beta_0 = \pi/4$. For activity number 0.001, the external concentration gradient has no effect on the concentration field. As the activity number is increased, the effect of the concentration gradient is seen. In Fig 3b, both the effects are of similar order, and in Fig 3c, the external concentration gradient dominates. Here, activity is taken as 1 on the red surface and 0 on the blue surface.

## 4. Translational and rotational velocity

The concentration field around the particle gives rise to a diffusio-osmotic flow in a thin region around the particle. This can be visualised as a slip velocity $\mathbf{u}_s$. This slip velocity is obtained from the concentration gradients as

$$\mathbf{u}_s = \mu_0 (\mathbf{I} - \mathbf{nn}) \cdot \nabla c \qquad (34)$$

Here $\mu_0$ represents the scaled mobility coefficient along the surface. For $\mu_0 > 0$, the relative interaction of solute molecules with respect to the solvent molecules is repulsive; for $\mu_0 < 0$, the relative interaction of solute molecules with respect to the solvent molecules is attractive. A Janus particle has different surface coverage on its two faces, which may alter the particle-solute interactions along the surface. We define the mobility to be uniform in each half as,

$$\mu_0 = \begin{cases} \mu_+ & 0 < \theta < \dfrac{\pi}{2} \\ \mu_- & \dfrac{\pi}{2} < \theta < \pi \end{cases} \quad (35)$$

The dimensionless slip velocity is calculated using (34) and the concentration field from (33) as

$$\mathbf{u}_s = \frac{\mu_0}{A+1}\left(\left(\frac{3A}{2}(-\sin\theta\cos\beta + \sin\phi\cos\theta\sin\beta) + \sum_{l=0}^{l=\infty}\frac{\alpha_l}{l+1}\frac{dP_l(\cos\theta)}{d\theta}\right)\mathbf{e}_\theta + \frac{3A}{2}\cos\phi\sin\beta\mathbf{e}_\phi\right) \quad (36)$$

There are two contributions to the slip velocity; the first arises from the external gradient, while the second arises due to the asymmetry in activity. Additionally, the slip velocity has components in both $\mathbf{e}_\theta$ and $\mathbf{e}_\phi$ directions.

## 4.1 Swimming velocity

The slip on the surface gives rise to a diffusiophoretic motion. It is determined using the Lorentz reciprocal theorem as

$$\mathbf{V}_{swim} \cdot \hat{\mathbf{f}}_i = -\int\int_S \mathbf{n} \cdot \boldsymbol{\sigma}_i \cdot \mathbf{u}_s dA.$$

Here, $\boldsymbol{\sigma}_i$ is the stress tensor associated with a point force $\hat{\mathbf{f}}_i$, $\mathbf{n}$ is the normal vector on the surface and $\mathbf{V}_{swim}$, the swimming velocity. For a unit point force, $\mathbf{n}\cdot\boldsymbol{\sigma}_i = 1/(4\pi)\hat{\mathbf{e}}_i$; the swimming velocity then becomes,

$$\mathbf{V}_{swim} = -\frac{1}{4\pi}\int\int_S \mathbf{u}_s dA.$$

Substituting, $\mathbf{u}_s$ from (36), we evaluate: (i) the swimming velocity due to external concentration gradient $\mathbf{V}_{grad}$, and (ii) due to self-diffusiophoresis of Janus particle $\mathbf{V}_{Jan}$. We write $\mathbf{V}_{swim}$ as

$$\mathbf{V}_{swim} = \mathbf{V}_{grad} + \mathbf{V}_{Jan} \quad (37)$$

here,

$$\mathbf{V}_{Jan} = -\frac{1}{4\pi(A+1)}\iint_S \mu_0 \sum_{l=0}^{l=\infty}\frac{\alpha_l}{l+1}\frac{dP_l(\cos\theta)}{d\theta}\mathbf{e}_\theta dA \quad (38)$$

Evaluating the above integral yields,

$$\mathbf{V}_{Jan} = \frac{1}{8(A+1)}(\alpha_+ - \alpha_-)(\mu_+ + \mu_-)\mathbf{e}_z$$

The limit $A \to 0$ translates to the case with no external concentration gradient. Here our result converges to that obtained by Golestanian et.al[3]. The expression for the swimming velocity shows that a difference in activities of the two surfaces is essential for a particle to propel itself in a uniform concentration.

We now evaluate the second term in (37). The swimming velocity due to external gradient is given by,

$$\mathbf{V}_{grad} = -\frac{3A}{8\pi(A+1)}\iint_s \mu_0((-\sin\theta\cos\beta + \sin\phi\cos\theta\sin\beta)\mathbf{e}_\theta + \cos\phi\sin\beta\mathbf{e}_\phi)dA \quad (39)$$

Here, $\mathbf{e}_\theta$ and $\mathbf{e}_\phi$ are unit vectors in the $\theta$ and $\phi$ direction. Converting to cartesian coordinates using :

$$\mathbf{e}_\theta = \cos\theta\cos\phi\mathbf{e}_x + \cos\theta\sin\phi\mathbf{e}_y - \sin\theta\mathbf{e}_z$$
$$\mathbf{e}_\phi = -\sin\phi\mathbf{e}_x + \cos\phi\mathbf{e}_y,$$

we obtain swimming velocity in $x, y$ and $z$ directions, respectively

$$\mathbf{V}_{grad,x} = \frac{-3A}{8\pi(A+1)}\int_0^\pi\int_0^{2\pi}\mu_0(-\sin\theta\cos\beta\cos\theta\cos\phi + \cos^2\theta\sin\phi\cos\phi\sin\beta - \cos\phi\sin\beta\sin\phi)r\sin\theta d\theta d\phi \quad (40)$$

$$\mathbf{V}_{grad,y} = \frac{-3A}{8\pi(A+1)}\int_0^\pi\int_0^{2\pi}\mu_0(-\sin\theta\cos\theta\sin\phi\cos\beta + \cos^2\theta\sin^2\phi\sin\beta + \sin\beta\cos^2\phi)r\sin\theta d\theta d\phi \quad (41)$$

$$\mathbf{V}_{grad,z} = \frac{3A}{8\pi(A+1)}\int_0^\pi\int_0^{2\pi}\mu_0(-\sin^2\theta\cos\beta + \sin\beta\sin\theta\cos\theta\sin\phi)r\sin\theta d\theta d\phi \quad (42)$$

Evaluating the above integrals, yields:

$$\mathbf{V}_{grad} = -\frac{A\sin\beta}{2(A+1)}(\mu_+ + \mu_-)\mathbf{e}_y - \frac{A\cos\beta}{2(A+1)}(\mu_+ + \mu_-)\mathbf{e}_z$$

Thus, the net swimming velocity is:

$$\mathbf{V}_{swim} = \frac{(\alpha_+ - \alpha_-)(\mu_+ + \mu_-)}{8(A+1)}\mathbf{e}_z - \frac{A(\mu_+ + \mu_-)(\sin\beta\mathbf{e}_y + \cos\beta\mathbf{e}_z)}{2(A+1)} \quad (43)$$

The unit vectors $\mathbf{e}_y$ and $\mathbf{e}_z$ are in the particle frame of reference, which rotates with the particle rotational velocity. The rotation of the particle changes the direction of the unit vectors with respect to the stationary reference frame. To account for this, we represent the velocity in terms of the stationary frame where the unit vectors are $\mathbf{e}_{y_0}$ and $\mathbf{e}_{z_0}$ as shown in Fig 1b. The transformation between these two reference frames is given by,

$$\mathbf{e}_y = \mathbf{e}_{y_0}\cos\theta_0 + \mathbf{e}_{z_0}\sin\theta_0$$
$$\mathbf{e}_z = -\mathbf{e}_{y_0}\sin\theta_0 + \mathbf{e}_{z_0}\cos\theta_0.$$

A substitution of the above transformations in (41) and using $\beta = \beta_0 + \theta_0$, results in the translational velocity in the stationary frame as,

$$\mathbf{V}_{swim,0} = \begin{pmatrix} -\frac{(\alpha_+ - \alpha_-)(\mu_+ + \mu_-)\sin\theta_0}{8(A+1)} - \frac{A(\mu_+ + \mu_-)\sin\beta_0}{2(A+1)} \end{pmatrix}\mathbf{e}_{y_0} \\ + \begin{pmatrix} \frac{(\alpha_+ - \alpha_-)(\mu_+ + \mu_-)\cos\theta_0}{8(A+1)} - \frac{A(\mu_+ + \mu_-)\cos\beta_0}{2(A+1)} \end{pmatrix}\mathbf{e}_{z_0} \quad (44)$$

here, $\theta_0$ is a function of time, given by $\theta_0 = \int_0^t \Omega_x dt$, with $\Omega_x$ as the angular velocity and $\beta_0$ is the constant angle between the concentration gradient and the stationary frame. We see that as A is increased (i.e. strength of applied concentration gradient increased), the relative contribution of self-diffusiophoresis to swimming velocity reduces.

## 4.2 Rotational velocity

The slip velocity on the surface (36) has both $\theta$ and $\phi$ components, which may induce a rotational velocity. Using Lorentz reciprocal theorem and following H Masoud and H Stone[22], we derive the rotational velocity:

$$\Omega = -\frac{3}{8\pi}\int_s \mathbf{n}\times\mathbf{u}_s dA \qquad (45)$$

Here, $\mathbf{n}$ is the surface normal vector $\mathbf{n} = \mathbf{e}_r$. The slip velocity (36) can be written as $\mathbf{u}_s = \mathbf{u}_{Jan} + \mathbf{u}_{grad}$, where $\mathbf{u}_{jan}$ is the contribution to slip from surface activity and $\mathbf{u}_{grad}$ is the contribution to slip from the external concentration gradient. We evaluate the rotational velocity from both the contributions separately. The slip velocity due to activity, $\mathbf{u}_{Jan}$, is along $\mathbf{e}_\theta$. Consequently the rotational velocity contribution from this has the direction $\mathbf{n}\times\mathbf{u}_{Jan} = \mathbf{e}_r\times\mathbf{e}_\theta = \mathbf{e}_\phi$. To evaluate the integral, $\mathbf{e}_\phi$ is converted to cartesian coordinates, $\mathbf{e}_\phi = -\mathbf{e}_x\sin\phi + \mathbf{e}_y\cos\phi$. This results in,

$$\int_s \mathbf{n}\times\mathbf{u}_{Jan} dA = \int_s \left(\frac{\mu_0}{A+1}\sum_{l=0}^{l=\infty}\frac{\alpha_l}{l+1}\frac{dP_l(\cos\theta)}{d\theta}\right)\left(-\mathbf{e}_x\sin\phi+\mathbf{e}_y\cos\phi\right)dA \qquad (46)$$

The first term is a function of $\theta$ alone, and the surface integral over $\sin\phi$ and $\cos\phi$ are 0. Therefore,

$$\int_s \mathbf{n}\times\mathbf{u}_{Jan} = 0$$

The rotational velocity contribution from self-diffusiophoresis is zero because of the axisymmetry of the Janus particle. To evaluate the rotational velocity contribution coming from the external concentration gradient, we substitute $\mathbf{u}_{grad}$ in (45). This yields,

$$\Omega = -\frac{3}{8\pi(A+1)}\int_s \mu_0 \frac{3A}{2}[(-\sin\theta\cos\beta+\sin\beta\sin\phi\cos\theta)\mathbf{e}_\phi + (\sin\beta\cos\phi)(-\mathbf{e}_\theta)]dA \qquad (47)$$

The unit vectors in spherical coordinates are transformed to cartesian coordinates as,

$$\mathbf{e}_\theta = \cos\theta\cos\phi\mathbf{e}_x + \cos\theta\sin\phi\mathbf{e}_y - \sin\theta\mathbf{e}_z$$
$$\mathbf{e}_\phi = -\sin\phi\mathbf{e}_x + \cos\phi\mathbf{e}_y$$

Using this, we obtain the $x, y$ and $z$ components of the rotational velocity.

$$\Omega_x = -\frac{9A}{16\pi(A+1)} \int_0^\pi \mu_0 \int_0^{2\pi} (\sin\theta\cos\beta\sin\phi - \sin\beta\cos\theta)\sin\theta d\phi d\theta$$

$$\Omega_y = -\frac{9A}{16\pi(A+1)} \int_0^\pi \mu_0 \int_0^{2\pi} \sin^2\theta\cos\beta\cos\phi d\phi d\theta \quad (48)$$

$$\Omega_z = -\frac{9A}{16\pi(A+1)} \int_0^\pi \mu_0 \int_0^{2\pi} \sin^2\theta\sin\beta\cos\phi d\phi d\theta$$

the integrals which determine $\Omega_y$ and $\Omega_z$ are zero as $\int_0^{2\pi} \cos\phi d\phi = 0$. The particle has an angular velocity only along $\boldsymbol{e_x}$. After evaluating the integral for $\Omega_x$, we obtain

$$\Omega = \frac{9A\sin\beta(\mu_+ - \mu_-)}{16(A+1)}\boldsymbol{e}_x \quad (49)$$

The slip on the surface due to the external concentration gradient breaks the axisymmetry, causing the particle to rotate. This shows that asymmetry in surface mobilities is essential for the particle to rotate.

### 4.3 Particle trajectory

The rotational and translational velocities are now expressed as,

$$\Omega_x = \frac{d\theta_0}{dt} \quad \text{and} \quad V_{swim,0} = \frac{d\boldsymbol{s}}{dt} \quad (50)$$

From Fig 1b, the relation between the angles in two frames of reference is

$$\frac{d\theta_0}{dt} = \frac{d\beta}{dt}$$

The displacement is represented as $\boldsymbol{s} = y_0\boldsymbol{e_{y_0}} + z_0\boldsymbol{e_{z_0}}$, where $(y_0, z_0)$ is the position of origin. Using (44) and (50), we obtain

$$\frac{dy_0}{dt} = -\frac{(\alpha_+ - \alpha_-)(\mu_+ + \mu_-)\sin\theta_0}{8(A+1)} - \frac{A(\mu_+ + \mu_-)\sin\beta_0}{2(A+1)}$$

$$\frac{dz_0}{dt} = \frac{(\alpha_+ - \alpha_-)(\mu_+ + \mu_-)\cos\theta_0}{8(A+1)} - \frac{A(\mu_+ + \mu_-)\cos\beta_0}{2(A+1)} \quad (51)$$

$$\frac{d\beta}{dt} = \frac{9A\sin\beta(\mu_+ - \mu_-)}{16(A+1)}$$

Integrating the above equation provides us the trajectory of the particle. The rotational velocity is integrated first as it is an independent equation.

$$\int_{\beta_0}^{\beta} \frac{d\beta}{\sin\beta} = \int_0^t \frac{9A(\mu_+ - \mu_-)}{16(A+1)} dt \; .$$

The above equation can be integrated analytically to get,

$$\tan\left(\frac{\beta}{2}\right) = \tan\left(\frac{\beta_0}{2}\right)\exp\left(\frac{9A(\mu_+ - \mu_-)t}{16(A+1)}\right). \tag{52}$$

To solve for net displacement, we substitute solution for $\beta$ given by (52) in (51) and obtain

$$y_0 = -\int_0^t \left(\frac{(\alpha_+ - \alpha_-)(\mu_+ + \mu_-)\sin\theta_0}{8(A+1)} - \frac{A(\mu_+ + \mu_-)\sin\beta_0}{2(A+1)}\right)dt \tag{53}$$

and

$$z_0 = \int_0^t \left(\frac{(\alpha_+ - \alpha_-)(\mu_+ + \mu_-)\cos\theta_0}{8(A+1)} - \frac{A(\mu_+ + \mu_-)\sin\beta_0}{2(A+1)}\right)dt \tag{54}$$

From Fig 1, $\theta_0(t) = \beta(t) - \beta_0$ and $\beta$ is obtained from (52). Using this, the above integration is performed computationally to find the trajectory of the Janus particle. In the next section, we see how the theoretical framework helps analyse the different trajectories and reorientation for different Activity numbers.

## 5. Results and discussion

So far the theoretical framework which helps obtain the rotational and the translational velocity of a Janus particle placed in an external concentration gradient has been established. We will now discuss the effects of different parameters Activity number, asymmetry in surface mobility and activity on the reorientation time and the trajectory of the particle. We define $\Delta\mu = \mu_+ - \mu_-$ and $\Delta\alpha = \alpha_+ - \alpha_-$ to represent the difference in mobilities and activities.

### 5.1 Particle re-orientation

A Janus particle exhibits both rotational and translational motion. We first discuss the rotioanl motion which helps the partilce orient itself along (either up or down) the concentration gradient. Using equation (52), we find the evolution of angular displacement with time for $\Delta\mu = \pm 1$ (Fig 4a) and for different activity numbers (Fig 4b). Equation (52) shows that the orientation follows an exponential decay or growth (depending upon the sign of $\Delta\mu$). We observe that as $t \to \infty$,

$$\text{For } \Delta\mu > 0, \quad \tan\left(\frac{\beta}{2}\right) \to \infty \quad \Rightarrow \quad \beta \to \pi$$

$$\text{For } \Delta\mu < 0, \quad \tan\left(\frac{\beta}{2}\right) \to 0 \quad \Rightarrow \quad \beta \to 0$$

The rotation of the particle for $\Delta\mu = \pm 1$ is shown in Fig 4a for $\beta_0 = \pi/4$. For $\Delta\mu > 0$, from (49) we see that the axis of rotation is along the $\boldsymbol{e}_x$ direction (i.e. rotation is clockwise). Whereas, for $\Delta\mu < 0$, the particle rotates along $-\boldsymbol{e}_x$ direction (rotation is anticlockwise). Fig 4b shows how the reorientaion time depends on the activity number A. As the activity number increases (i.e. relative strength of applied concentration gradient increases), the particle re-orients faster.

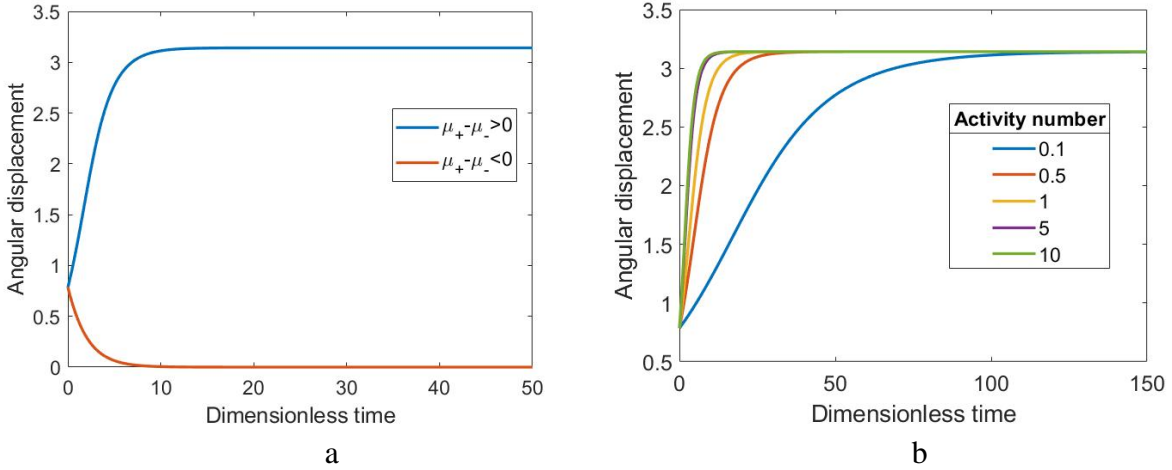

Figure 4: Angular displacement as a function of time for $\beta_0 = \pi/4$. (a) The particle rotates clockwise if the net solute-particle interaction is repulsive and rotates anticlockwise if the net interaction is attractive. Here A=10. (b) The angular displacement as a function of time for different activity numbers. The angular displacement tends to $\pi$ because the interactions are taken as repulsive, $\Delta\mu = +1$.

Quantifying the re-orientation time has implications on optimizing the design of microfluidic experiments. Hence we now obtain expressions for this and analyze how it is effected by *the activity number A*. The re-orientation time is defined as the time required for 99% orientation. The angular displacement is given by (52) as

$$\tan\left(\frac{\beta}{2}\right) = \tan\left(\frac{\beta_0}{2}\right)\exp\left(\frac{9A\Delta\mu t}{16(A+1)}\right)$$

Expressing 99% orientation as $\hat{\beta}$ and reorientation time as $t_{0.99}$. For $\Delta\mu = +1$, $\hat{\beta} = 0.99\pi$; while for $\Delta\mu = -1$, $\hat{\beta} = 0.01$.

$$t_{0.99} = \frac{16}{9\Delta\mu}\ln\left(\frac{\tan(\frac{\beta}{2})}{\tan\left(\frac{\beta_0}{2}\right)}\right)\left(1+\frac{1}{A}\right). \tag{55}$$

From (55) we observe that as the activity number is increased, the reorientation time reduces, saturating to a value depending upon the initial orientation $\beta_0$. Interestingly, for a fixed activity number, a Janus particle with a larger value of $\beta_0$ takes less time to reorient, as the rotational velocity is higher for a particle with higher initial angle. To obtain more physical insights, we look at the dimensional reorientation time.

We convert (54) to dimensional form using the timescale, $t_{ch} = a^*/\gamma^*\mu^*$, here $a^*$ is the radius of the particle, $\gamma^*$ is the applied concentration gradient and $\mu^*$ is the characteristic mobility coefficient. Using the definition of activity number, and representing the dimensional reorientation time as $t_{0.99}^*$, we get

$$t^*_{0.99} = \frac{16a^* \times 10^{-3}}{9\mu^*(\Delta\mu)} \ln\left(\frac{\tan\left(\frac{\beta}{2}\right)}{\tan\left(\frac{\beta_0}{2}\right)}\right)\left(\frac{|\alpha^*| \times 10^{-3}}{D^*\gamma^{*2}N_A^2} + \frac{1}{\gamma^* N_A}\right) \tag{56}$$

For a 10μm particle in water, with oxygen as the solute, $D^* = 2.3 \times 10^{-9} m^2 s^{-1}$, $\mu^* = 8.87 \times 10^{-33} m^5 s^{-1}$, $|\alpha^*| \sim O(10^{19})m^{-2}s^{-1}$. Substituting $\Delta\mu = 1$ and $\beta_0 = \pi/4$ in (56) we obtain

$$t^*_{0.99} = \frac{120.98}{\gamma^{*2}} + \frac{16.75}{\gamma^*} \tag{57}$$

We now compare the reorientation time (57) with the rotational Brownian time scale $t^*_{Brown}$. The latter scales as[23] $t^*_{Brown} \sim \frac{8\pi\eta^* a^{*3}}{k_B T^*}$ here η is the viscosity of the fluid, '$a^*$' is the radius of the particle and $k_B T^*$ is the thermal energy. For a particle size of 10μm in water at 298K, $t^*_{Brown} \sim O(10^3)$ sec. For a 10μm particle in water with oxygen as the solute, the dependence of reorientation time on concentration gradient is shown in Fig 5b. When the reorientation time is larger compared to $t^*_{Brown}$, the rotational Brownian noise keeps changing the direction of motion hindering the reorientation of the particle. The direction of motion is randomized under these conditions and the particle shows a noisy or random walk. This situation prevails to the left of the dashed vertical line in Fig. 5b. On the other hand, when the reorientation time is smaller compared to $t^*_{Brown}$, rotational Brownian noise is still present, but the particle reorients and moves along/opposite to the gradient. We divide the graph (Fig 5b) into two regions: i) where the Brownian noise has a significant role and the particle re-orientation is hindered (on the left of dashed line shown in Fig 5b) and ii) where the Brownian noise will have a negligible effect on the reorientation (on the right of the dashed line shown in Fig 5b). In Fig 5b, the nature of the dependency on concentration gradient changes from $O(1/\gamma^*)$ for low concentration to $O(1/\gamma^{*2})$ when the concentration gradient is increased. The change in functional form in the two regions is due to change in the effect which is dominant.

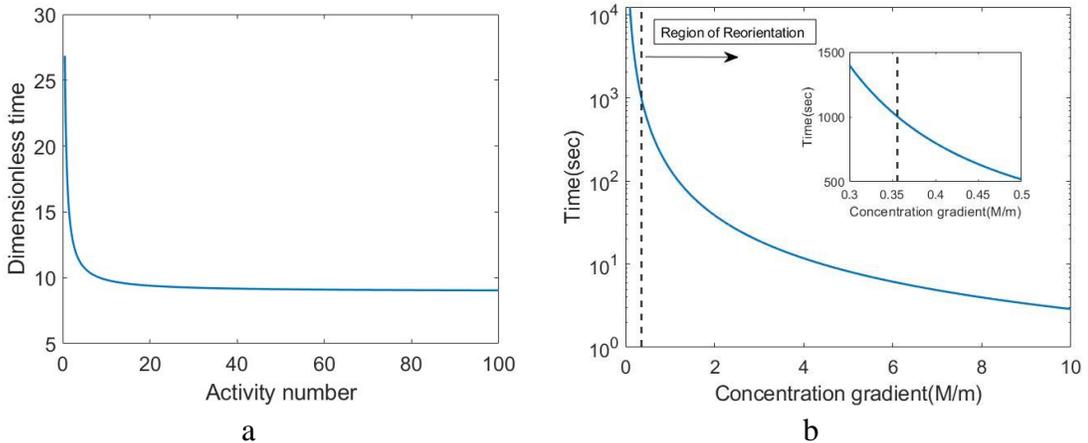

Figure 5: (a) $t_{0.99}$ as a function of activity number. The dimensionless reorientation time reduces as the activity

number is increased, finally saturating to a value of 8.95 for $\beta_0 = \pi/4$. (b) shows the dimensional reorientation time for a 10μm particle placed in water with oxygen as the solute. Inset shows zoomed in graph at low concentration gradient.

## 5.2 Particle trajectory

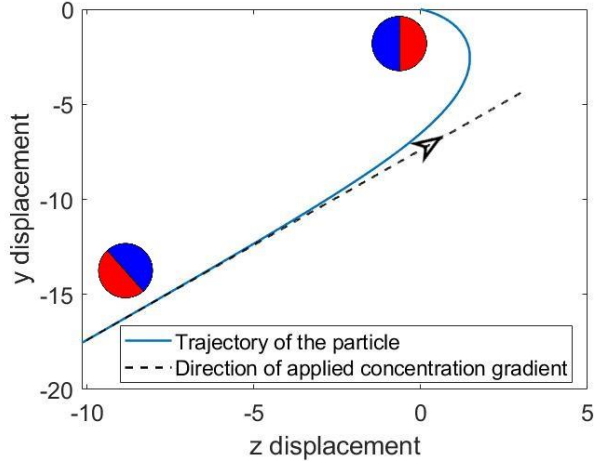

Figure 6: Trajectory of a Janus particle. Starting from the origin, the particle eventually moves along the concentration gradient towards the low concentration region for A=0.1, $\Delta\mu = +1, \Delta\alpha = +1$. Concentration increases along the arrow. The axis of self-propulsion is initially horizontal, later aligns opposite to the concentration gradient ($\beta = \pi$).

Having discussed the reorientation time we now focus on the translational motion which determines the particle trajectory. In Fig. 6, we plot the trajectory of the particle when an external concentration gradient is imposed. For the case depicted the interactions are taken as repulsive ($\Delta\mu > 0$) and the particle rotates clockwise. Consequently the face with higher repulsive interaction faces the lower concentration region, minimizing the energy of the system. Due to repulsive interactions, the particle moves away from the higher concentration region and $\beta \to \pi$.

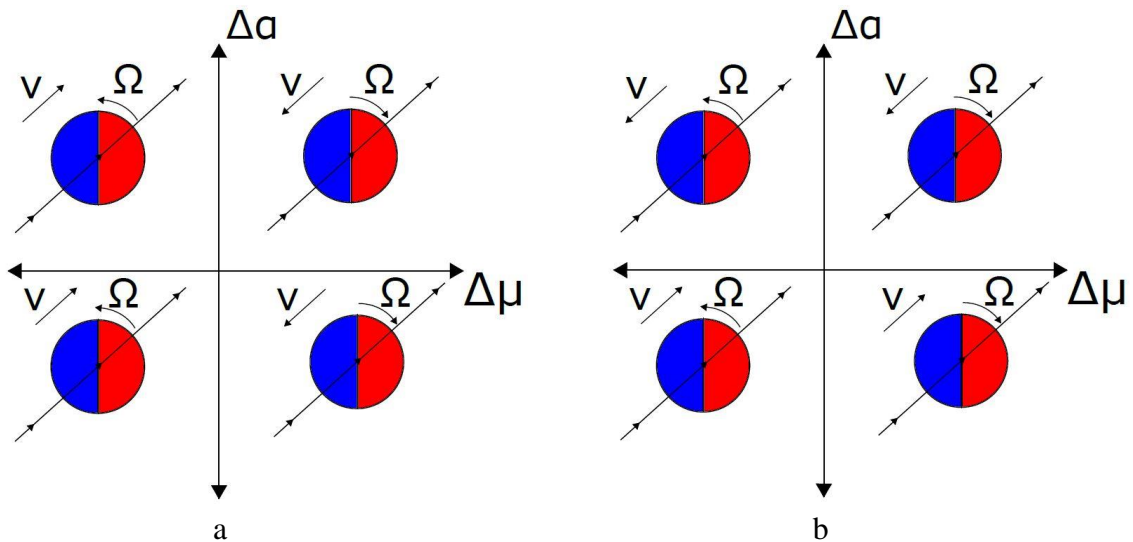

Figure 7: Classification of the parameter space where the particle shows different directions of translation and rotation for a) low A(local gradient dominates) and b) high A(external gradient dominates). Here, $\Delta\alpha = \alpha_+ - \alpha_-$ and $\Delta\mu = $

$\mu_+ - \mu_-$. When $\Delta\mu$ and $\Delta\alpha$ are of opposite sign(quadrant II,IV), the two effects compete, which leads to change in direction of swimming depending on which of the two effects are dominant.

The trajectory of the particle depends both on the asymmetry in surface mobility($\Delta\mu$) and activity($\Delta\alpha$). The parameter combination of surface mobility and activity determines the particle behavior. Based on the sign of $\Delta\alpha$ and $\Delta\mu$, there are four possible cases. All the cases are qualitatively captured in a "phase diagram" with $\Delta\mu$ and $\Delta\alpha$ as the parameters (as depicted in Fig7). The diagram shows the direction of translation and rotation for each case. When $\mu_+ + \mu_- = 0$, the particle merely rotates without translation, this has been excluded in the phase diagram. In the first quadrant($\Delta\mu > 0, \Delta\alpha > 0$) the particle rotates clockwise. With $\Delta\alpha > 0$, the self-generated concentration gradient and the external concentration gradient are in the same direction. In this case, the external gradient enhances the swimming velocity. Similarly in third quadrant ($\Delta\mu < 0, \Delta\alpha < 0$), both the gradients are in the same direction, again enhancing the swimming velocity. Wheras, in the second quadrant($\Delta\mu < 0, \Delta\alpha > 0$), the particle rotates anti-clockwise. A positive value of $\Delta\alpha$ in this case, creates a local concentration gradient which acts opposite to the external gradient. This leads to a competetion between the two gradients. The direction of swimming in this case depends on the relative strength of the two gradients (shown in Fig7a and 7b). Similarly, in the fourth quadrant ($\Delta\mu > 0, \Delta\alpha < 0$), the two gradients are in oppsite direction, leading to competition between the two. Here again the relative strengths of the two gradients determines the particle trajectory.

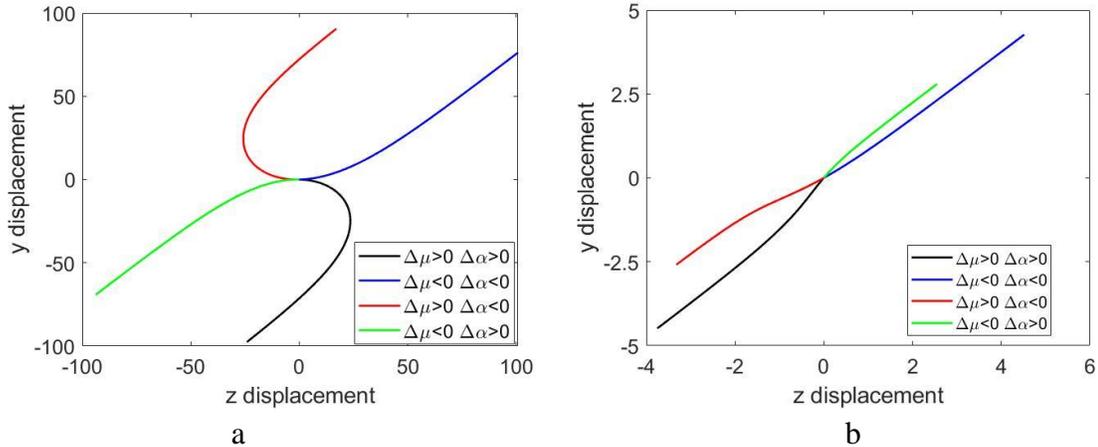

Figure 8: Trajectory of a Janus particle starting from the origin placed in linear concentration gradient for a)A=0.01 and b)A=1. Swimming direction reverses if $\Delta\mu$ and $\Delta\alpha$ have opposite sign.

Fig 8 shows the trajectory of a Janus particle for different combinations of $\Delta\mu$ and $\Delta\alpha$. We first consider the case $\Delta\mu\Delta\alpha > 0$ ( both are positive or negative). The corresponding trajectories are shown by blue and black curves in Fig 8. Here, the external and self-generated concentration gradients are in the same direction. In Fig 8, the time of integration is same for all trajectories. The black and blue trajectories are longer than the green and red. This is due to the enhancement in swimming velocity when $\Delta\mu\Delta\alpha > 0$. Comparing Fig 8a and 8b we see that the swimming direction for black and blue trajectory remains same because both the concentration gradient are in the same direction. The swimming direction in this case is independent of the activity number. Whereas, when $\Delta\mu\Delta\alpha < 0$ (shown by green and red trajectories), the two effects oppose each other . Consequently, as the activity number increases (relative strength of imposed concentration gradient increases), the swimming direction reverses.

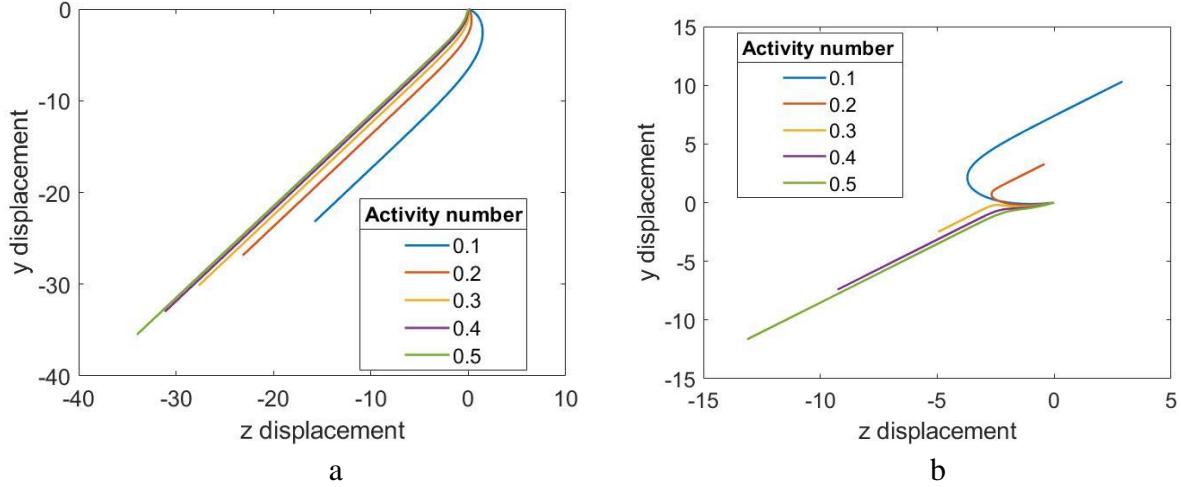

a                                    b

Figure 9: Trajectory of a Janus particle starting from the origin for the two cases a) $\Delta\mu > 0, \Delta\alpha > 0$ and b) $\Delta\mu > 0, \Delta\alpha < 0$ with $\beta_0 = \pi/4$. In Fig 9a, both the effects are in the same direction, and the direction of swimming is independent of activity number. In Fig 9b, the two effects compete, and a reversal in swimming direction is observed.

To study the effect of activity number on the trajectory, we plot the trajectory at different activity numbers. Fig 9a shows the trajectories for $\Delta\mu\Delta\alpha > 0$ at different activity numbers. As the activity number increases the particle travels a shorter distance before orienting itself along the concentration gradient. In this case, the external concentration gradient, increases the swimming velocity. This is reflected in Fig 9a, where the particle travels a longer distance as the activity number increases .Fig 9b shows the trajectories for $\Delta\mu\Delta\alpha < 0$ (i.e. quadrant II and IV) at different activity numbers. In this case, the swimming direction reverses as the activity number increases from 0.1 to 0.5. The reversal in swimming direction takes place at a critical activity number, $A_{cr}$. The direction of the external gradient is exactly opposite to that of the self-generated gradient only when the particle has reoriented. At critical activity number, the velocity of the particle is zero. Substituting $\beta = 0$ in (43) and equating the swimming velocity to zero yields,

$$A_{cr} = \frac{\Delta\alpha}{4} \qquad (58)$$

The direction of the trajectories shown in Fig 9b change across this critical number ($\Delta\alpha = 1, A_{cr} = 0.25$).

## 6. Conclusions

In this work, we investigate the behavior of a Janus particle under the influence of an externally imposed linear concentration gradient of a non electrolytic solute. Exploiting the characteristic time scales of the system the governing equations are simplified to a linear system of equations. This enables us to obtain an analytical solution which gives insights into system behavior. The Lorenz reciprocal theorem is used to compute the slip velocity and the swimming velocity of the particle.

We showed that the $e_\phi$ component of the slip velocity causes the particle to rotate leading to its reorientation. Our approach clearly shows that symmetry breaking in both the surface activity and surface mobility is essential for the particle to reorient and move along the concentration gradient. The direction of rotation depends on the relative interaction of the two faces with the solute molecules. The reoreintation is such that the face with a relatively less repulsive interaction with the solute molecules faces the higher concentration region ; as this minimizes the energy of the system. As a consequence of this reorientation, the self-generated local concentration gradient can either be along or opposite to the external concentration gradient. When the local concentration gradient is against the external concentration gradient ($\Delta\mu.\Delta\alpha < 0$), the direction of swimming depends on the relative strengths of the two effects. We also calculate the critical activity number at which the direction of swimming reverses. On the other hand, when the local concentration gradient is along the external concentration gradient, the external concentration gradient enhances the net swimming velocity. This can be elegantly depicted by dividing the parameter space of $\Delta\mu$ and $\Delta\alpha$ into four quadrants and qualitatively showing the the direction of swimming and rotation in each of them.

Furthermore, we showed that as the activity number increases, the reorientation time (dimensionless) continuously decreases saturating to a value depending upon the initial orientation of the Janus particle to the external concentration gradient. We also compare the effect of Brownian noise on the reorientation by comparing the respective timescales. The current analysis is valid for concentration gradients when we can neglect the role of Brownian noise.

Current work focuses on a half faced Janus particle (surface coverage $\eta = \pi/2$). This can be extended to account for an arbitrary coverage $\eta$. Translational velocity of the particle is written as, $V_{swim} = V_{Janus} + V_{grad}$. These components are given by,

$$V_{Janus} = \frac{-1}{(A+1)D} \sum_{l=0}^{l=\infty} \left(\frac{l+1}{2l+3}\right) \alpha_{l+1} \left(\frac{\mu_l}{2l+1} - \frac{\mu_{l+2}}{2l+5}\right) \hat{e}_z \tag{59}$$

$$V_{grad,z} = \frac{-A\cos\beta}{4(A+1)} \left[2(\mu_+ + \mu_-) + (\mu_+ - \mu_-)(\cos^3\eta - 3\cos\eta)\right] \hat{e}_z \tag{60}$$

$$V_{grad,y} = \frac{-A\sin\beta}{8(A+1)} \left[4(\mu_+ + \mu_-) - (\mu_+ - \mu_-)(\cos^3\eta + 3\cos\eta)\right] \hat{e}_y. \tag{61}$$

Due to symmetry, there is no contribution from $V_{grad, x}$ (see eq. 40). The rotational velocity of the particle for arbritrary coverage is given by

$$\Omega_x = \frac{9A\sin\beta\sin^2\eta}{16(A+1)}(\mu_+ - \mu_-). \tag{62}$$

We refer the readers to the Appendix for a detailed derivation. The above equation shows that changing the coverage on the particle will change the magnitude of the translational and rotational velocity. Thus, the trajectory of these particles can be expected to be qualitatively similar to the those shown in Fig 8 and 9.

The assumption of vanishingly small Peclet number neglects the effect of advection on the particle trajectory. Advective effects weakens the concentration gradient leading to a lower slip velocity, resulting in a low diffusiophoretic velocity[19]. Specifically, the velocity reduces as $O(Pe^2)$ for weak advective effects. However, at higher Peclet numbers, the coupling between the solute and momentum transport can lead to non-intuitive results, such as a maxima in translational velocity with increase Peclet number[24]. However, these modifications do not effect the direction of translation. The rotational velocity is also expected to reduce in the presence of advective effects, as it reduces the magnitude of diffusio-osmotic slip[19]. However, a detailed analysis is needed for an in-depth understanding of the effect of solute advection on the trajectory of the Janus particles in the presence of an externally applied concentration gradient.

The current framework can be extended for weak non-linear concentration gradients, by expanding the concentration field around the particle in Taylor series and retaining the first order terms. For a decaying concentration field from a source present at a given site, the gradient also decays as we move away from the site. Therefore, we can calculate a crtical distance between the Janus particle and the site, beyond which the particle will not sense the chemical gradient and reorient along the concentration gradient. A Janus particle inside this critical distance will reorient and move towards/away from the site. This can help design microfluidic devices to carry out anti-susceptibiliy test of bacteria.

## Appendix

Here we provide the detailed derivation for the expressions (59)-(62) which describe the translational and rotational velocities for a particle with an arbitrary coverage $\eta$. Here the mobility coefficient and activity are given by

$$\alpha, \mu = \begin{cases} \alpha_+, \mu_+ & 0 < \theta < \eta \\ \alpha_-, \mu_- & \eta < \theta < \pi \end{cases}. \tag{A1}$$

Expressing surface activity and mobility using Legendre polynomials as a basis, we obtain $\alpha = \sum_{l=0}^{l=\infty} \alpha_l P_l(cos\theta)$ and $\mu = \sum_{l=0}^{l=\infty} \mu_l P_l(cos\theta)$. The concentration field and the slip velocity are obtained using these coefficients ($\alpha_l, \mu_l$) in (33) and (36) respectively. The modified velocity expressions are obtained from (38), (41), (42), and (48). Using properties of Legendre polynomials, the contribution due to activity is found to be

$$V_{Janus} = \frac{-1}{(A+1)D} \sum_{l=0}^{l=\infty} \left(\frac{l+1}{2l+3}\right)\alpha_{l+1}\left(\frac{\mu_l}{2l+1} - \frac{\mu_{l+2}}{2l+5}\right) e_z \tag{A2}$$

Evaluating integral (41) for arbitrary coverages, we obtain

$$V_{grad,y} = \frac{-3A\sin\beta}{8(A+1)}\left[\mu_+ \int_0^{\eta}(\cos^2\theta+1)\sin\theta d\theta + \mu_- \int_{\eta}^{\pi}(\cos^2\theta+1)\sin\theta d\theta\right] e_y. \tag{A3}$$

The above expression simplifies to

$$V_{grad,y} = \frac{-A\sin\beta}{8(A+1)}\left[4(\mu_+ + \mu_-) - (\mu_+ - \mu_-)(\cos^3\eta + 3\cos\eta)\right]e_y. \quad (A4)$$

Similarly, $V_{grad,z}$ is given by

$$V_{grad,z} = \frac{-3A\cos\beta}{4(A+1)}\left[\mu_+\int_0^\eta \sin^3\theta d\theta + \mu_-\int_\eta^\pi \sin^3\theta d\theta\right]e_z \quad (A5)$$

This simplifies to

$$V_{grad,z} = \frac{-A\cos\beta}{4(A+1)}\left[2(\mu_+ + \mu_-) + (\mu_+ - \mu_-)(\cos^3\eta - 3\cos\eta)\right]e_z. \quad (A6)$$

The slip velocity contribution due to activity is along $e_\theta$, and therefore it does not contribute to particle rotation. Evaluating (48) with mobility coefficient defined as (A1), we obtain

$$\Omega_x = \frac{9A\sin\beta\sin^2\eta}{16(A+1)}(\mu_+ - \mu_-) \quad (A7)$$

# DATA AVAILABILITY

Data sharing is not applicable to this article as no new data were created or analyzed in this study